# Onboarding in Open Source Software Projects: A Preliminary Analysis

Fabian Fagerholm*, Patrik Johnson*, Alejandro Sánchez Guinea*, Jay Borenstein†, and Jürgen Münch*

*Department of Computer Science
University of Helsinki
P.O. Box 68 (Gustaf Hällströmin katu 2b)
FI-00014, Finland
Email: {firstname.lastname}@cs.helsinki.fi

†Department of Computer Science
Stanford University
353 Serra Mall
Stanford, CA 94305 USA

†Facebook
1601 Willow Road
Menlo Park,
CA 94025 USA
Email: borenstein@cs.stanford.edu

*Abstract*—Nowadays, many software projects are partially or completely open-source based. There is an increasing need for companies to participate in open-source software (OSS) projects, e.g., in order to benefit from open source ecosystems. OSS projects introduce particular challenges that have to be understood in order to gain the benefits. One such challenge is getting newcomers onboard into the projects effectively. Similar challenges may be present in other self-organised, virtual team environments. In this paper we present preliminary observations and results of in-progress research that studies the process of onboarding into virtual OSS teams. The study is based on a program created and conceived at Stanford University in conjunction with Facebook's Education Modernization program. It involves the collaboration of more than a dozen international universities and nine open source projects. More than 120 students participated in 2013. The students have been introduced to and supported by mentors experienced in the participating OSS projects. Our findings indicate that mentoring is an important factor for effective onboarding in OSS projects, promoting cohesion within distributed teams and maintaining an appropriate pace.

*Keywords*—*onboarding; open source software projects; virtual teams; mentoring; global software development; distributed software development; case study*

## I. INTRODUCTION

For years, software companies around the world have engaged with Open Source Software (OSS) projects and communities. The motivation for participating in, supporting, or actually establishing open source projects can stem from a desire to reduce development costs and increase the levels of innovation [1]. OSS development is similar to Global Software Development (GSD) in many ways and projects are often highly decentralised with participants from a wide range of geographical locations and cultural backgrounds.

In many application domains, engagement with OSS is an inevitable part of the computing business. In order to participate in existing software ecosystems (such as Android or the LAMP stack), companies may need to adopt open source development and licensing approaches in order to combine their offering with the infrastructure provided by the ecosystem. Another reason to participate in open source projects is to be able to conduct projects of a size that exceeds the capabilities of an individual organisation. In addition, engagement in open source projects can serve as a recruitment strategy that allows companies to analyse the performance and talents of different developers around the world without the need to set up special infrastructure for that purpose [2]. OSS also plays an important role in government IT and the open source approach is often considered an enabler for technology and knowledge transfer to developing countries (e.g. [3], [4]). The Open Source development approach has also greatly influenced software businesses using other kinds of licensing models [5].

Despite the benefits that OSS projects can offer for companies, actually obtaining those benefits requires understanding and managing a number of challenges. However, guidance to address these challenges is widely missing in many areas [6]. The most important among such challenges probably relates to the nature of the development process followed in many OSS projects.

Open source developers have a large degree of freedom to decide how to manage themselves within their projects. OSS development teams can be characterised as *self-organised virtual teams*. We refer to "virtual teams" as "teams whose members use technology to varying degrees in working across locational, temporal, and relational boundaries to accomplish an interdependent task" [7]. Self-organisation in open source projects refers to the lack of formally appointed leaders or indications of rank or role, and to a large degree of shared power [8]. Such teams accomplish organisation of their work through self-assignment and "soft delegation" where participants ask each other to perform tasks rather than command or direct task assignments.

Because of the need to quickly get involved in OSS projects, an important question is how to support the entry of new members into OSS projects. *Onboarding*, or *organisational socialisation*, is "the process that helps new employees learn the knowledge, skills, and behaviours they need to succeed in their new organizations" [9]. Studies on onboarding have examined different essential aspects such as mentoring [10], virtual teams [11], and other factors that impact the process [12], [13]. Despite the importance attributed to onboarding, only few studies have directly examined onboarding in open source projects, but some work does exist (e.g. [14], [15]). However, factors that have been identified as highly important in previous research on onboarding have been largely neglected. Onboarding in open source projects is still poorly understood: little research on social processes related to team maintenance has been conducted [6], and as far as we can see, the gap still exists. Due to their similarities, results from studying onboarding in OSS projects may also be applicable to GSD projects.

In this paper, we present a preliminary analysis of a study on onboarding in open source projects. The ultimate aim is to derive guidelines for onboarding in different kinds of environments, while this paper describes initial results. The rest of this paper is organised as follows. In Section II, we describe the international collaborative program which forms the context for this study. We then present the preliminary research design, hypotheses, and method in Section III. In Section IV, we present preliminary results from the study. Finally, we discuss the results and future research in Section VI.

## II. Collaboration Program

The context for our study is a large collaborative program with several participating open source communities, companies, and universities. The project was established and coordinated through Facebook's Education Modernization program. While the program itself was educational in nature, its other properties make it suitable for examining onboarding in an open source environment. In this study, we focus on a subset of the program and examine onboarding in four open source projects. In this section, we describe the program with particular emphasis on aspects that are relevant for this study.

The program was initially piloted by Jay Borenstein, Education Modernizer at Facebook and lecturer in Computer Science at Stanford University. The pilot was conducted through a collaboration between a Stanford University course and the Phabricator Open Source Project and its principal developer, Evan Priestley. Based on the success of the initial pilot, the program was expanded substantially in 2013. The international reach of Facebook's Education Modernization program was leveraged to introduce the element of cross university collaboration. More than a dozen universities were part of the collaboration in 2013.

Each university integrated the program into their computer science curriculum in the way that was suited to their academic calendar and infrastructure. For example, at the Department of Computer Science, University of Helsinki, students joined the program through the Software Factory laboratory for research and education [16], [17].

Nine open source projects (see Table I) were carefully selected by Mr. Borenstein to participate in the critical mentoring role for this collaborative program. The main criteria for being a mentor in this program is the capacity to be responsive to students, overall quality as a software engineer, and a high activity level in the open source project itself. A total of approximately 130 students participated, referred to below as developers. In this study, a subset of these developers are compared against developers in the same OSS projects who did not receive the onboarding support provided by the collaborative program.

Prior to the January start of the program, developers submitted their OSS project preferences through their respective faculty members to Mr. Borenstein. They were then assigned to virtual teams with 4–8 (median: 5) participants from at least two different universities. Each OSS project had two separate virtual teams working for it, and provided an experienced senior developer to mentor the developer teams.

Two activities supported onboarding in the projects: a kickoff activity, and continuous mentoring during the project. At the start in January 2013, developers gathered for a three-day kickoff Hackathon event at Facebook's Menlo Park headquarters in California, USA. This component of the program is noteworthy in that it takes significant resources to bring involved faculty members, OSS mentors, and students together to a single location. Facebook's Education Modernization team handled all financial, logistic, and coordination-related aspects of the event. We emphasise that only the student developers participated in this event; other developers already participating in the same OSS projects were not present.

During the event, developers got acquainted with other members of their virtual teams and met their open source mentors. The mentors provided hands-on, practical training to the developers, so that they could gain the basic technical skills required to participate in and contribute to their projects.

After the kickoff session, developers returned to their home universities and continued work as virtual teams. A set of practices were suggested for the mentors, but each mentor was free to apply them in their own way. In particular, the practices of conducting daily meetings, being available via email or chat on a daily basis, and setting a clear, high-level goal after a period of familiarisation were suggested. The mentors continued to support the teams during the remainder of the project by, e.g., participating in online forum and mailing list discussions, conducting or participating in online meetings through video conferencing and chat, helping developers find and understand tasks, reviewing code contributions and giving feedback on them, and helping to coordinate work through issue tracking systems.

Developers were free to work on any tasks relevant to their projects. Initially, mentors would typically direct developers to small tasks suitable for novices, assuming that as the developers became more proficient, they would begin to take the initiative and tackle tasks of greater complexity. Most tasks were programming tasks of different kinds, ranging from small bug-fixes to complicated new features. Other tasks included writing test cases, creating new issues in tracking systems when new bugs were found, and improving some non-functional aspect of the software, taking into account maintainability, performance, and user experience.

The developers were integrated into each open source project and community by the regular procedures of that community. They have thus been exposed to the regular norms and implicit policies of each community. In addition, they have received support from their mentors, from their local and remote team members, as well as any support provided by their home universities. We consider the context suitable for examining the onboarding process and the effect of these additional support structures on the process. In company settings, it appears realistic that similar support structures can be enacted both to enable entry into external open source projects as well as to enable third parties to enter projects driven by the company itself.

## III. Preliminary Research Study

The goal of this study is to characterise and understand the onboarding process in the OSS projects under study, with particular consideration of the role of mentoring as part of that process. Our study focuses on the following four open source projects: Kotlin, Phabricator, Ruby on Rails, and Socket.IO (see Table I). We chose to examine these projects because the student developers enrolled in the Software Factory Project course at the Department of Computer Science, University of Helsinki, were members of the virtual teams working on these projects. By examining these projects first, we were able to observe them more closely and make appropriate choices in the research design. The long-term goal is to develop guidelines for onboarding in open source projects, but at this stage, preliminary results of characterisation and understanding are the initial focus. In this section, we describe the study as far as necessary in order to understand the preliminary results.

TABLE I. PARTICIPATING OPEN SOURCE PROJECTS.

| Project name | URL | Included in this study? |
|---|---|---|
| Freeseer | http://freeseer.github.io/ | No |
| Kotlin | http://kotlin.jetbrains.org/ | Yes |
| MongoDB | http://www.mongodb.org/ | No |
| Mozilla OpenBadges | http://openbadges.org/ | No |
| ReviewBoard | http://www.reviewboard.org/ | No |
| Phabricator | http://phabricator.org/ | Yes |
| PouchDB | http://pouchdb.com/ | No |
| Ruby on Rails | http://rubyonrails.org/ | Yes |
| Socket.IO | http://socket.io/ | Yes |

As noted, open source teams can be characterised as self-organised virtual teams. Such teams are also possible in other kinds of GSD settings. These teams are not formed once and for all, but are in a continuous state of change. The formation strategy requires members to engage with each other, as they cannot rely on management to assign members to their teams. New members often need to learn technical and social skills as they enter the project. Personal motivation may play a significant role in the ability to enter projects and remain an active contributor.

We examine how certain supporting activities affect onboarding and thus the productivity and communication activity among team members. We hypothesise that factors such as face-to-face, workshop-like meetings, interaction with co-located participants, and mentoring can serve to support onboarding by increasing the chance that participants are exposed to, select, and perform tasks in the project on their own.

One way of analysing the issue described above is to look at the initiation of developers in open source projects, comparing developers who receive specific onboarding support with developers who enter the projects in the "usual" way – without onboarding support.

*A. Research Design*

The study follows a mostly quantitative multiple-case study approach [18]. We replicate the same study across several projects in order to increase the generalisability of the results. Following the Goal-Question-Metric approach [19], [20], we have operationalised the goals and questions of the study into quantitative metrics. Analysing the metrics will provide information to answer the questions and thus address the overall goal of the study. In addition, qualitative observations made during the study will be used to help interpret the metrics and the contribution of each question to the overall goal. The GQM measurement goal is shown in tabular form in Table II. In this study, we focus mainly on the following questions:

TABLE II. DEFINITION OF GQM MEASUREMENT GOAL.

| Object | Onboarding process |
|---|---|
| Purpose | Characterisation and Understanding |
| Focus | Contribution level over time |
| Viewpoint / Stakeholder | Project manager / Open source mentor |
| Context | International collaboration project (see main text) |

Q1 How much time does a developer communicate in the project over time?
Q2 How much time and effort does a developer put into the project over time?
Q3 How much does a developer contribute to the project over time?
Q4 How much mentoring does each mentor give to their respective team(s)?
Q5 What aspects of onboarding are considered important by project mentors and developers?
Q6 What is the influence of mentoring on the performance of the developers?

*B. Metrics*

We derived several metrics from measurement goals following the GQM approach [19], [20]. Among these, most of our attention in this preliminary stage has been put on *activity*, due to its intuitive role as an objective indicator of onboarding. We define activity as a compound of metrics that together measure the effective participation of a developer in a particular project.

Fritz et al. present and use a metric for onboarding, comprising the interest, knowledge and interaction of developers [21]. Inspired by this work, we define activity as a linear combination that considers number of commits, number of pull requests and number of interactions. All these direct metrics correspond to data that can be gathered from the public GitHub revision control system that is used by all OSS projects considered.

A commit is understood in the context of revision control systems as representing submission of the latest changes in the source code to the repository. A pull request in turn is characterized by GitHub as a notification that is made to others about changes pushed to the repository, so that interested parties can review changes, discuss potential modifications and push follow-up commits if necessary [22]. An interaction is defined as a single message posted in a GitHub discussion forum. These discussions may be attached to commits or pull requests, and may concern potential changes to the source code, messages justifying changes, and general messages related to the development of a pull request or commit. Each such message is counted as one interaction.

A refinement of the coefficients of the linear combination for the activity measure is expected as the study continues. Also, more parameters will probably be added to include the knowledge of developers with respect to their projects, which has currently been left out of consideration in this preliminary work.

*C. Sample*

For the purpose of comparing between mentored and non-mentored developers, we sampled two groups among developers. To form the group of mentored developers, we took a random sample of 20 developers involved in the collaboration project. For the non-mentored group, we took a random sample of the same size of developers who have contributed to the same OSS projects as the developers of the mentored group, but who have not been involved in the collaboration project.

TABLE III. TOTAL ACCUMULATED ACTIVITY.

| Developers | Commits | Pull requests | Interactions | Activity |
|---|---|---|---|---|
| Mentored | 64 | 111 | 251 | **426** |
| Non-mentored | 19 | 37 | 92 | **148** |

For the purpose of comparison over time, we defined a time series sampling strategy for obtaining the accumulated activity over time. We sampled the activity for each developer every week for the first 12 weeks of the project involvement. The definition of the first week required some assumptions to be made regarding the non-mentored group. For the mentored group, we simply took the initial week of the collaboration project as week 1. For the non-mentored group, we defined the initial week as the week previous to the date when the first activity count was found for each developer. The weeks are thus relative to when the developer began their onboarding process.

## IV. PRELIMINARY RESULTS

We obtained both the total accumulated activity and the accumulated activity over time for both groups. Table III shows the absolute numbers for each direct metric and the compound activity metric. Developers in the mentored group show roughly three times more activity than the non-mentored group.

While the activity metric cannot completely answer the questions described in Section III-A, it contributes to answering questions 1, 2, 3, and 6. In Table III, we can see that mentored developers interact more than non-mentored developers (Q1), presumably spending more time on the project (Q2). Mentored developers also produce more commits and pull requests, which indicates that they may contribute more to the project (Q3).

The accumulated activity over time is shown in Figure 1. The numbers in the plot have been scaled to the range 0–10, in order to make the projects comparable. The figure depicts the data collected for the mentored and non-mentored groups with the corresponding linear regression line that helps to see the growth pattern more clearly. As can be seen in the figure, the activity of developers that have received onboarding support has grown considerably more with time compared to those that did not receive such support. This indicates that onboarding support has a big impact in the projects under study (Q6).

Closer examination of Figure 1 shows that initially, activity in both groups is roughly the same. At week three, however, the supported group dramatically increases its activity, which then levels off for another three weeks. After that, the activity again increases, and after a plateau, starts increasing steadily. There are some fluctuations in the non-supported group, but the changes in activity level are smaller. At this stage of the study, we do not have a well-grounded explanation for why the increases in activity are timed at approximately three-week intervals. While it may be an anomaly, other possible reasons could be that task size or effort tends to converge so that tasks generally require three weeks to be completed, or that there is something inherent in human information processing or communication abilities that sets the pace of understanding among software developers in distributed projects to progress in approximately three-week intervals. However, we stress that these are merely guesses.

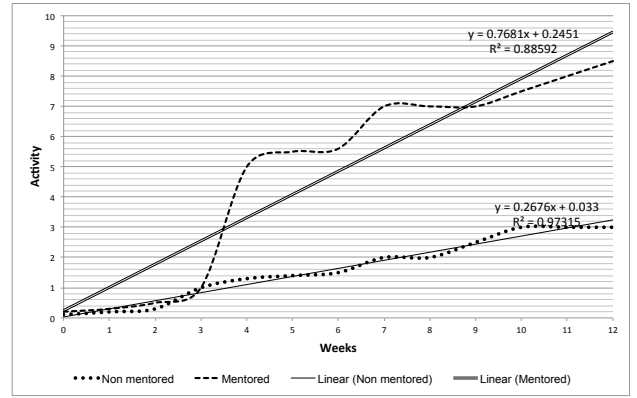

Fig. 1. Cumulative activity of supported and non-supported developers over time.

In summary, the results indicate that onboarding support has a large influence on activity in the projects under study. The data analysed here, while incomplete, indicates that the onboarding process may benefit from the involvement of a mentor. It would be interesting to observe if the growth in activity will continue being larger even when mentoring has stopped, for those that have received it – in other words, whether mentoring has a permanent effect on developer performance. This however, cannot be supported by solid evidence at this stage of the project. For such purpose the collaboration project will have to come to its end and our study continued, in order to observe the developers that were involved. Thus, an interesting path of research can be drawn by following up this study.

## V. LIMITATIONS

Onboarding is a complex construct with multiple impact factors and context factors that may limit the applicability of the results in other cases. Being a preliminary analysis, we have not yet examined the exact factors in detail. Based on the data analysed in this study, we cannot distinguish between mentoring and other onboarding support factors. Other contextual factors may play a role. For example, developers in the group receiving onboarding support may be more involved, i.e., spend more time, more regularly, with the project. Also, the kickoff event may be another factor which impacts the results.

In this preliminary analysis, we included only part of the data available. We focused on four of the nine projects, and thus the results may be biased towards the particular characteristics of these projects. Expansion of the sample is necessary to confirm that the results apply to all the projects in the open source collaboration program.

The observed effect may stem from factors that are either not present in other cases, or that are moderated by other factors not present in this study. For example, the use of student subjects could cause the study to be less applicable among professional developers, either because of different levels of expertise, because of differing organisational constraints, or because of different reward mechanisms (salary versus credit points).

A further limitation is that we have not examined whether developers in the non-supported group has received any treatment during the observed period of time that would impact the results. We can thus not be entirely sure what kinds of factors influence their activity. However, we expect our sample of developers to represent the variety of conditions in which open source developers in the projects under study normally work. Furthermore, even though the exact conditions of this group is unknown, our results do indicate that the onboarding support given to the project teams has an effect on activity.

We have not statistically evaluated the construct validity of our activity metric. Since this paper presents preliminary results, it can be considered a pilot study which contributes to the evaluation of construct validity. There are theoretical grounds to expect that supporting onboarding by different means and in different stages of involvement will shorten the time required for developers to become effectively onboard in an organisation. Furthermore, there are theoretical grounds to expect that individuals who have progressed further in their onboarding process will display higher productivity than other individuals. Our activity metric appears to correlate with these theoretical assumptions. Still, no single study can prove construct validity, and further work is needed to evaluate the validity of our activity metric.

One possible limitation of this study is that subjects may have altered their behaviour in response to the study or to the fact that they are being assessed as students by their universities. While this possible limitation needs to be taken into account, we note that i) the subjects did not receive information about their performance from the researchers, ii) the educational assessment of the subjects was not performed by the researchers, and iii) developers in companies are also frequently evaluated. We thus argue that any biasing effect of the study itself on the onboarding process is likely to be extremely low.

## VI. CONCLUSION

In this study, we examined the onboarding process of a subset of developers engaged in a collaborative project with several participating open source communities, companies, and universities. The aim of the study was to characterise and understand the onboarding process in these projects. The study is a preliminary analysis which we expect to lead to guidelines for onboarding in open source projects.

In the study, we used the GQM approach to derive an activity metric, which measures the amount of commits, pull requests, and interactions in online discussion fora that a developer has produced during a specific period of time. We compared the cumulative activity of developers who received onboarding support as part of the collaborative project with that of developers working in the same open source projects, but who did not receive the onboarding support. In our preliminary results, we found that the developers in the first group showed more activity during the first 12 weeks of their participation than developers in the second group.

Our interpretation of the results is that the onboarding support given to the first group of developers has influenced their onboarding process, allowing them to become more active at an earlier stage. We thus find support for the hypothesis that onboarding support increases the chance of developers to be exposed to, select, and perform tasks in a proactive and self-directed manner in open source projects. We assume that this applies in other kinds of GSD settings where the organisational and team structure is similar.

Among the factors involved in onboarding support, which include an initial co-located kickoff session, interaction with co-located participants, and mentoring, we assume that the last factor is highly important for maintaining developers' motivation and for attaining good cohesion within virtual teams, since it involves expert guidance that is sustained during a long period. It thus has the potential to be one of the largest factors that influence onboarding. We also assume that it helps maintain clear objectives in open source projects, which directs the efforts of developers to work collaboratively on mutual, interdependent goals.

This study has opened a number of questions to be addressed in future work. First, this paper examined only one metric, i.e., activity. Even our preliminary characterisation of onboarding (see Section III-A) includes several GQM questions with other metrics. A more comprehensive set of metrics promises to measure onboarding better.

Another question is whether the effect of onboarding support is permanent even after some or all of the support activities are removed. Future work could assess to what degree the effect is present, taking into account that the organisational context will change since the universities involved will no longer reward credit points to the students involved. However, if onboarding has been very successful, students may continue working for their open source projects as volunteers.

The construct validity of both the activity metric and the entire set of metrics in our GQM graph should be evaluated. In future work, validity may be evaluated by eliciting expert assessment as to how far developers have come in their onboarding process. Also, an estimation of the difference in time spent by each developer on project-related tasks would allow us to reduce some of the identified limitations. Separating the effect of different components of the onboarding support activities can allow us to assess the relative importance of mentoring for onboarding. Expansion of the the sample will increase the validity of the results. Finally, taking other impact and context factors into account, we can begin to assess the applicability of the construct in different kinds of environments.

Our ultimate goal is to provide guidelines for companies wishing to accelerate onboarding in open source projects of importance to them. We expect that some companies are looking for guidelines on how to involve their own employees into external open source projects, while other companies are more interested in guidelines on how to accelerate onboarding of external contributors to open source projects conducted by them companies themselves. Some of the guidelines are likely applicable in both cases, while other guidelines are specific to one of the scenarios. For example, mentoring is likely to be applicable in both cases, while arranging Hackathon events or other kinds of co-located workshops may be more realistic in the second case. With appropriate modification, the guidelines may also be useful in other kinds of GSD settings.